\documentclass[12pt]{iopart}

\usepackage{iopams}
\usepackage{graphicx}
\usepackage[center]{subfigure}
\begin{document}

\title{Nonlinear light propagation in cholesteric liquid crystals with a
helical Bragg microstructure}

\author{Yikun Liu$^{1}$, Shenhe Fu$^{1}$, Xing Zhu$^{1}$, Xiangsheng Xie$%
^{1} $, Mingneng Feng$^{1}$, Jianying Zhou$^{1\ast}$, Yongyao Li$^{2}$, Ying
Xiang$^{1,3}$, Boris A. Malomed$^{4}$, Gershon Kurizki$^{5}$}

\address{$^{1}$State Key Laboratory of Optoelectronic Materials and Technologies, Sun
Yat-Sen University, \\Guangzhou 510275, China\\
$^{2}$Department of Applied Physics, South China Agricultural University,\\ Guangzhou 510642, China\\
$^{3}$Information Engineering College, Guangdong University of Technology,
Guangzhou, China\\
$^{4}$Department of Physical Electronics, School of Electrical Engineering,
\\Faculty of Engineering, Tel Aviv University, Tel Aviv 69978, Israel\\
$^{5}$Chemical Physics Department, Weizmann Institute of Science, Rehovot
76100, Israel\\}
\ead{stszjy@mail.sysu.edu.cn}
\vspace{10pt}
\begin{indented}
\item[]December 2014
\end{indented}

\begin{abstract}
Nonlinear optical propagation in cholesteric liquid crystals (CLC) with a
spatially periodic helical molecular structure is studied experimentally and
modeled numerically. This periodic structure can be seen as a Bragg grating
with a propagation stopband for circularly polarized light. The CLC
nonlinearity can be strengthened by adding absorption dye, thus reducing the
nonlinear intensity threshold and the necessary propagation length. As the
input power increases, a blue shift of the stopband is induced by the
self-defocusing nonlinearity, leading to a substantial enhancement of the
transmission and spreading of the beam. With further increase of the input
power, the self-defocusing nonlinearity saturates, and the beam propagates
as in the linear-diffraction regime. A system of nonlinear couple-mode
equations is used to describe the propagation of the beam. Numerical results
agree well with the experiment findings, suggesting that modulation of
intensity and spatial profile of the beam can be achieved simultaneously
under low input intensities in a compact CLC-based micro-device.
\end{abstract}

%
%
%
%
%

\section{Introduction}

The propagation of light waves at frequencies near the propagation
bandgap of nonlinear photonic structures is the subject of very broad
interest \cite{1,2,3,4,5,6}.  Most experiments in this field
were performed in fiber Bragg gratings\ \cite{15,16}, or other solid-state
materials, which may be silicon-on-isolator and AlGaAs \cite{27,28}. Many
applications rely on the use of nonlinear-optical effects, such as
bistability \cite{7}, compression and shaping of laser pulses \cite{8,9},
generation of gap solitons \cite{10,11,12}, storage and buffering of ultrashort
optical pulses \cite{13,13b}, ultrafast optical switches \cite{14}, etc.%
\newline
\indent In the spatial domain, Bragg solitons in continuous optical media
were predicted in various settings, that can be implemented using planar waveguides
and photonic crystals, but they have not yet been demonstrated experimentally. On the other hand,
discrete gap solitons were created in arrays of waveguides with the
self-focusing Kerr nonlinearity \cite{Silberberg}.

\indent In this work, we study the nonlinear beam propagation in a dye-doped
cholesteric liquid crystal (CLC), whose molecules are oriented so as to form
a periodic helical order in the longitudinal direction \cite{17}, thus
building a Bragg structure which affects circularly polarized light and
creates a steep propagation stopband, with the spectral width close to $80$
nm.An appropriate dopant added to the CLC matrix can reduce the nonlinear
threshold by several orders of magnitude due to the additional dye torque
\cite{J,YRShen,M,18,19,20}. Thus, the strong nonlinearity gives rise to a
low operational power, and requires shorter propagation lengths. These
advantages provide simultaneous modulation of the beam's intensity and
spatial structure at low input intensities in a compact sample. \newline
\indent We experimentally test the beam propagation in the CLC under
different input powers. To this end, the light beam is selected with the
carrier frequency located near the edge of the gap. A substantial increase
of the transmission is observed with the increase of the input power, due to
the blue shift of the propagation stopband induced by the self-defocusing
nonlinearity. The output spot size is also measured, showing spreading of
the beam under the action of the self-defocusing nonlinearity. With the
further increase of the input power, the self-defocusing nonlinearity
saturates, and finally leads to the propagation in the effectively
linear-diffraction regime. Thus, CLC settings offer the advantage of 
implementing the simultaneous modulation to the beam's profile and intensity 
by a single compact device. To model the experimental setting, we use
nonlinear couple-mode equations for the light propagation in the grating.
The numerical results agree well with the experiment findings.

\section{The cholesteric liquid crystals used in the experiment}
\indent The CLC medium, built of rod-like molecules, was produced by adding
a chiral agent to nematic liquid crystal 5CB ($n{_{0}}=1.53$, $n{_{e}}=1.72$%
, $dn=0.19$) supplied by Merck. The nematic phase exists in this
material at temperatures from $18^{\circ }$ C to $35^{\circ }$ C,
self-assembling to build a stacked periodic corkscrew-like structure, due to
the helical twist introduced by the chiral agent. The average refractive
index of the medium is $1.635$, and its birefringence is $0.19$. The pitch
of the CLC structure can be controlled by the concentration of the chiral
agent. Absorption dye is additionally doped into CLC to provide the dye
torque and thus enhance the third-order nonlinearity.

\indent The sample was fabricated with the thickness of 50 $\mathrm{\mu }$m,
and the density of the chiral agent was fixed at $17.4\%$, to induce the
propagation stopband of the spectral width $80$ nm, with the center set at $%
442$ nm, and the red band edge at $468$ nm. The CLC was doped by the DCM
[4-(Dicyanomethylene)-2-methyl-6-(4-dimethylaminostyryl)-4\textit{H}-pyran]
laser dye with the absorption peak at $470$ nm and the concentration of $1$
wt$\%$. Thus, suitable absorption can be achieved at the red edge of the
stopband. The samples were sandwiched by two Indium Tin Oxide (ITO) glass
substrates with a $50$ $\mathrm{\mu }$m thick spacer, coated by polyimide
and rubbed unidirectionally to impose the planar alignment. The CLC sample
was pulled into the cell by the capillary force. Molecules in the CLC are
helically aligned in the same direction in the plane parallel to the cell
surface. The scheme of the CLC sample is shown in Fig. 1.
\begin{figure}[tbp]
\centering{\includegraphics[width=6.5cm, height=6.5cm]{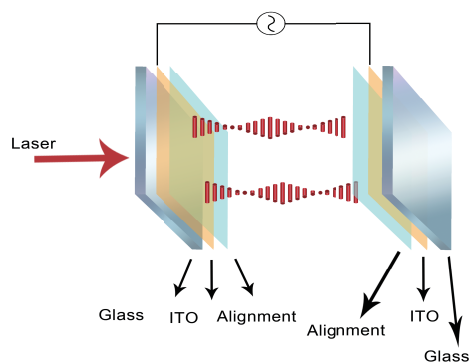}}
\centering{\caption{(Color online) A scheme of the setup.}
\label{fig_1}}
\end{figure}
\newline
\indent Previous studies successfully analyzed photo-tunable characteristics
of dye-doped CLCs \cite{21,22,apex}. Studies of nonlinear optical properties of
CLCs in the direction perpendicular to the helix were reported before \cite%
{b1,c1}. The spatial properties of the nonlinear beam propagation were
previously investigated in the direction perpendicular to the helix \cite{23}%
, which is  determined by the spatial discrete diffraction (due to the coupling between the two adjacent waveguides) and nonlinear effect.In these previous work the linear and nonlinear properties of CLC were discussed, and nonlinear spatial propagation properties and spatial soliton was investigated in the direction perpendicular to the helix\cite{Fra}. In the direction parallel to the helix, nonlinear effects have been already investigated\cite{Dwei1,Dwei2}. In this work, nonlinear spatial propagation in direction parallel to the helix will be investigated.

\section{The experiment}
\indent The experimental setup for demonstrating the stopband shift is shown in
Fig. 2(a). A xenon lamp was used as a white-light source with the circular
light polarization to measure the reflection spectrum of the CLC. The pump
is provided by an optical parametric oscillator operating at $468$ nm, which
is driven by a continuous mode-locked Ti:sapphire femtosecond laser with
repetition rate $80$ MHz and pulse duration $100$ fs. Here the femtosecond
pulse is used for generating the nonlinearity. The pump field with wavelength 
at $468$ nm is chosen for two reasons. First, the wavelength corresponds to the absorption peak 
of the dopant DCM, which gives rise to an additional dye torque acting on CLC molecules; 
second, such a wavelength is close to the red edge of the CLC stopband, which helps to 
observe a substantial increase of the transmission. The probe
beam was shone normally to the CLC cell. \newline
\indent Another setup for measuring the beam-propagation profile is shown in
Fig. 2(b). A single femtosecond probe beam was used for this purpose. The
spatial evolution was evaluated by measuring the beam's waist after passing $%
50$ $\mathrm{\mu }$m in the sample before reaching a CCD detector, starting
with a tightly focused input.


\begin{figure}[h]
\centering%
{\subfigure[]{\label{fig_2_a}\includegraphics[width=9cm,
height=4.5cm]{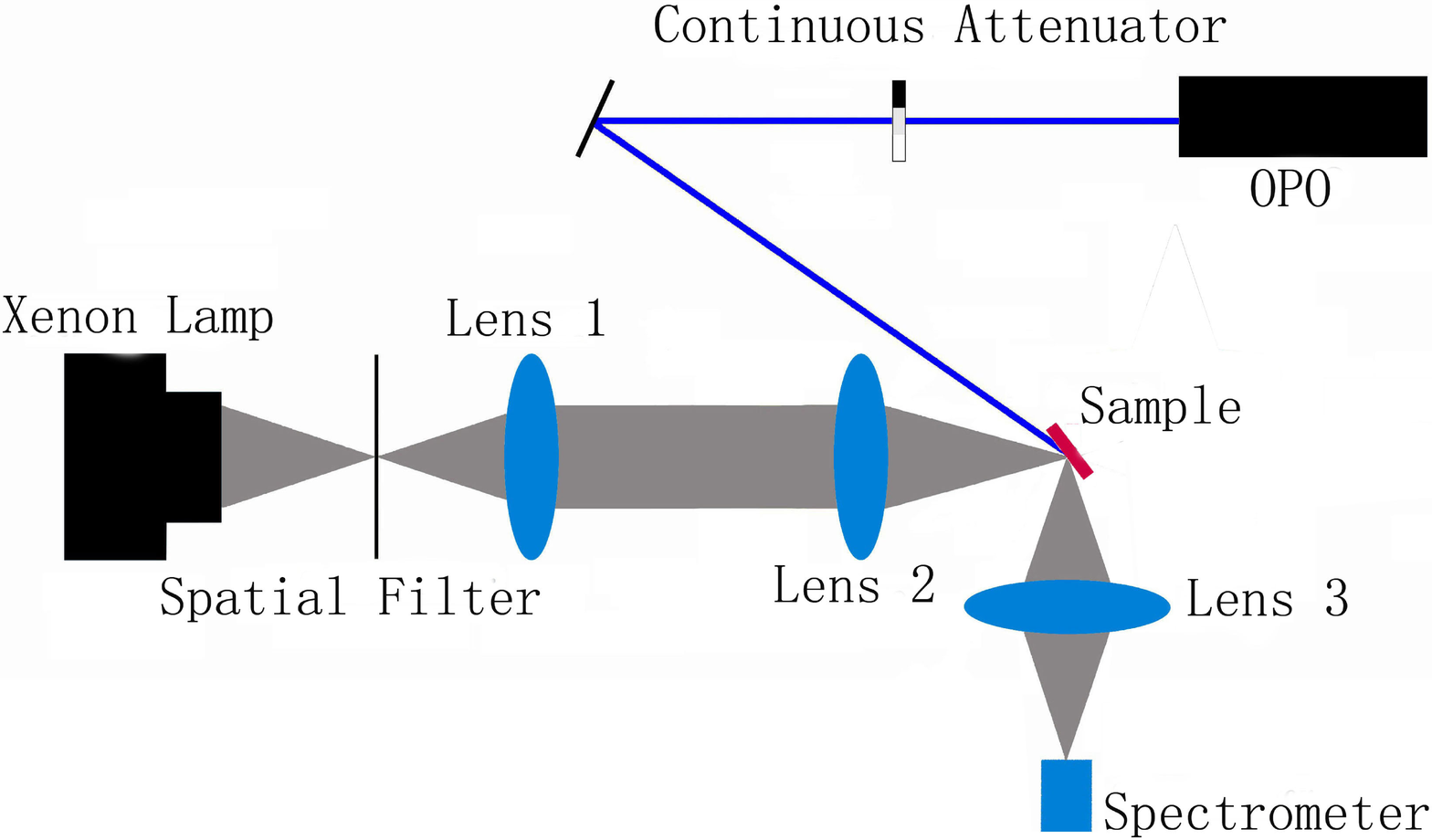}} \subfigure[]{\label{fig_2_b}%
\includegraphics[width=11cm, height=3.5cm]{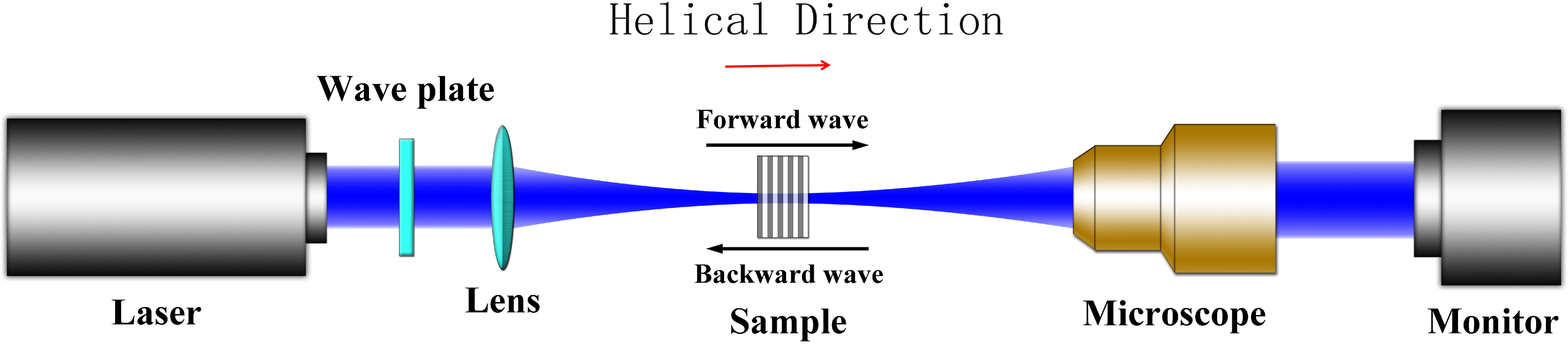}}
\caption{(Color online) Schematics of the experimental setups for measuring
the reflection spectra under pumping (a), and the transmission profile
(b). In the latter case, a single pump beam is used.}
\label{fig_2}}
\end{figure}


\indent As mentioned above, in our experiment the width of the
stopband is $80$ nm, and its edge is steep enough to generate a substantial
variation in the reflection/transmission due to a small shift of the
stopband. At an average pump power of $85$ mW (the corresponding peak-power
density is $1.25$ MW/cm$^{2}$), blue shift of the central wavelength of
stopband is observed (Fig. 3). According to the Bragg formula, the central wavelength
of the propagation stopband is $\lambda _{B}=2n_{\mathrm{eff}}d$, where $d$
and $n_{\mathrm{eff}}$ are the structural period and average refractive
index of the CLC. The bandwidth of the stopband, $\Delta \lambda $, is
determined by $\Delta \lambda =2dn\cdot d$,where $dn$ is the refractive
index modulation in the CLC. In the experiment, only the shift of the
stopband center is observed, without a change of the bandwidth. This result
suggests that the change of the average refractive index is the main
mechanism underlying the stopband shift.

The change of the refractive index in liquid crystals can be attributed to
various causes, such as the ultrafast Kerr effect \cite{liyan}, the thermal
\cite{KhooIC1} and photorefractive \cite{KhooIC2, lyK} effects, and the dye
torque \cite{J}. According to Ref. \cite{liyan}, in an undoped CLC sample,
optical intensity higher by a factor of $10^{3}$ was needed to produce the
same shift of the stopband as in the DCM-doped sample. This means the DCM
dopant plays a crucially important role for the shift.

To examine the contribution of the thermal effect to the CLC nonlinearity,
two additional experiments have been carried out. (1) A continuous-wave,
non-mode-locked laser with the same average power as that of the mode-locked
femtosecond laser, was applied to illuminate the sample. In this case, no
nonlinear shift of the stopband was observed, thus demonstrating that an
intensity-dependent nonlinear effect determines the shift, rather than the
average power that would be related to a thermally-induced effect.
(2) A pumping femtosecond pulse train with 2 kHz chopping, for which
the thermal effect should be much lower than in the case of a
continuous pulse, was also used to measure the stopband change. The
result is compared to that without chopping, showing that the
nonlinear shift of the bandgap does not depend on the chopper's
rate. This observation further suggests that the thermal effect
plays a negligible role in generating the nonlinear-optical effect.
The result obtained with the continuous wave suggests that the
photorefractive effect is not significant either for generating the
stopband shift.

Thus, the optical torque is a plausible mechanism which underlies the
nonlinear effect in the CLC. As a strong pulse is applied to the medium, an
additional dye torque is generated, leading to rotation of the CLC
molecules, and changing the refractive index. This outcome is not produced
by thermo-optical effects \cite{KhooIC1} induced by laser heating. Instead, the dye
torque is generated by electronically excited dye molecules, which is known
as the \textit{Janossy effect}. \cite{J,YRShen,M,18,19,20}.

The blue shift of the stopband indicates that the average
refractive index is reduced by the high-intensity pump, which may be
construed as a self-defocusing Kerr effect. In the initial situation without
a pump pulse, the rod-like molecules are oriented perpendicular to the helix
direction, which corresponds to the maximum refractive index. Under the
action of the pump beam, the molecules rotate out of cell plane, which leads to a decreased refractive index.This means without the excitation of pumping light, the molecules orientate perpendicular to the helix and paralleled to the polarization of pumping light, which give the refractive index $n_e$. As the sample are pumped by femtosecond pulse,the pumping light excite the dye, the dye will exert a dye-torque to reorient the LC molecules with angle ¦È with respect to cell plane,which result in the effective diffractive index $n_{eff}(\theta)$, obviously, $n_{eff} (\theta)<n_{e}$. In the experiment, we always observed the blue shifting of the stop band, and never observe red shifting, this result confirm the conclusion of rotate of LC out of the plane. This dye-torque should be asscribed to the Jannossy effect, as ref. 42 indicated that for pure CLC the pumping intensity should be about 1000 times higher.
 Due to the saturable absorption by the dopant dye,
the additional dye torque saturates too, eventually leading to the
saturation of the stopband shift. Accordingly, the inset in Fig. 3 shows
that the shift increases with the increase of the pump power, and then
reaches saturation when the pump power exceeds $0.7$ MW/cm$^{2}$). \newline
\indent 

\begin{figure}
\begin{center}
\includegraphics[width=5.5cm,height=3.8cm]{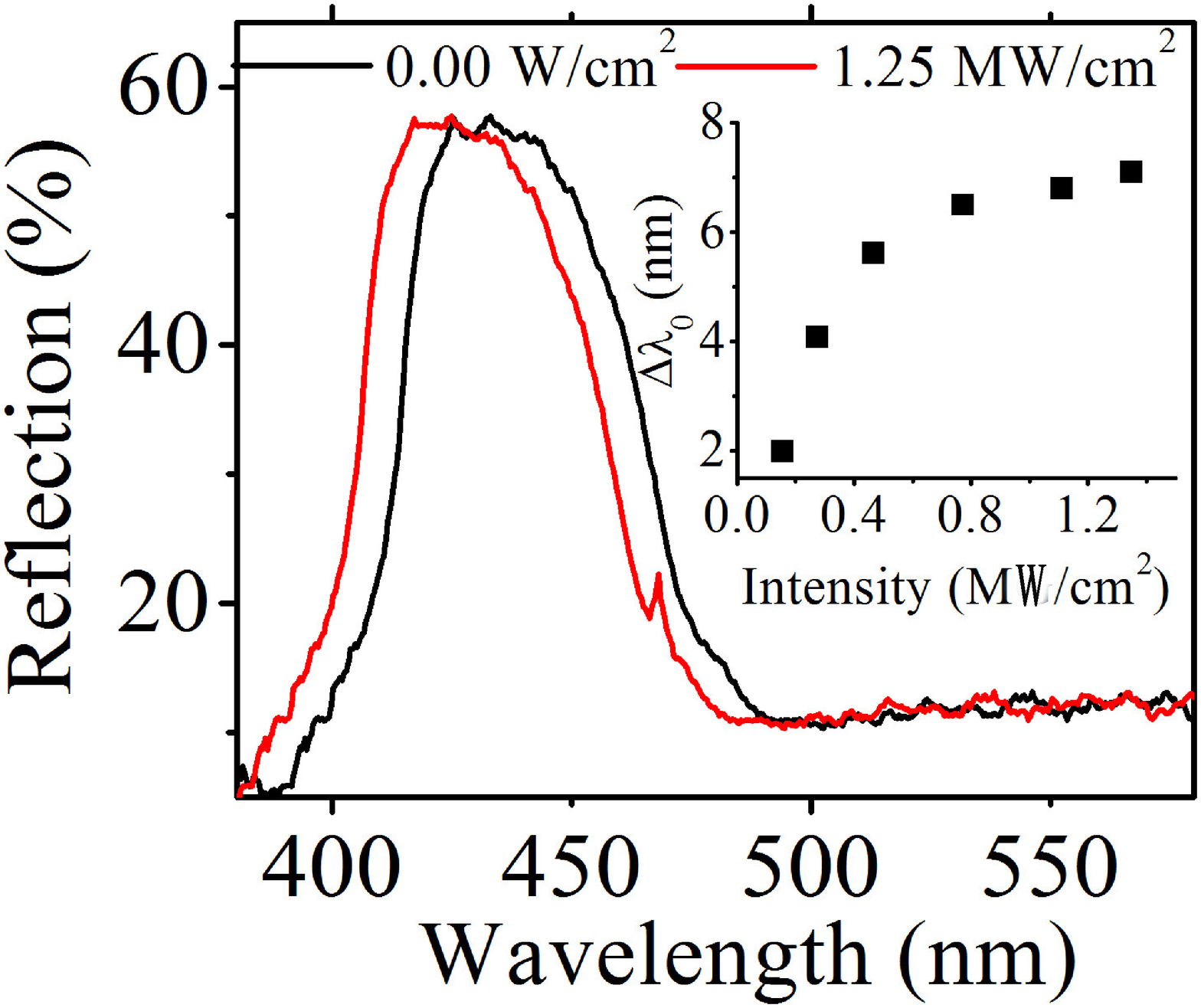}
\end{center}
\caption{(Color online) The reflection spectra produced by the CLC sample.
The photonic bandgap can be shifted under the action of the pump. The inset shows
experimental data for the shift of the bandgap as a function of the pump
power.}
\label{Fig. 3}
\end{figure}


\indent An obvious difference from the fiber Bragg gratings is that the
transverse field distribution in the CLC is not limited by the cross section
of the fiber, hence various phenomena in the transverse directions may be
observed. The propagation properties of the spatial beams were studied near
the edge of the stopband. A relation between the transmission coefficient
(normalized to that at the maximum of the transmitted power without the
Bragg stopband), for light at wavelengths $468$ nm and $492$ nm, and the
input intensity is shown in Fig. 4(a). The former wavelength is located at
the edge of the propagation stopband, while the latter one is located
outside. The results show that, due to the blue shift of the stopband, the
transmission at $468$ nm increases with the input intensity. On the other
hand, at $492$ nm, i.e., away from the stopband, the change of the
transmission is much less significant compared to the change of the
transmission at $469$ nm, which can be up to seven times higher. A $25\%$
change of the transmission may be explained as resulting from the saturable
absorption by the doping dye solvent, as well as from the possible variation
of the Fabry-Perot interference. For a high input intensity of $1.2$ MW/cm$%
^{2}$, the transmission at $468$ nm is approaching its value in the bulk
sample, suggesting that the stopband effect is completely eliminated by the
nonlinearity. In principle, the effective power-induced stopband suppression
may facilitate the generation of Bragg solitons \cite{10}.\newline
\indent The output profile of the beam was measured at $468$ nm, see Fig.
4(b). At low input intensities, the light is Bragg-reflected backward, hence
the transmission beam is not observed. As the input intensity increases, the
self-defocusing nonlinearity gives rise to the blue shift of the photonic
stopband, which allows the input beam to propagate through the sample with
strong transverse divergence. When the intensity increases further, the
self-defocusing nonlinearity saturates due to the above-mentioned absorption
saturation, in which case the beam propagation features the normal linear
diffraction.


\begin{figure}[h]
\centering%
\subfigure[]{\label{fig_4_a}\includegraphics[width=5.5cm,
height=3.8cm]{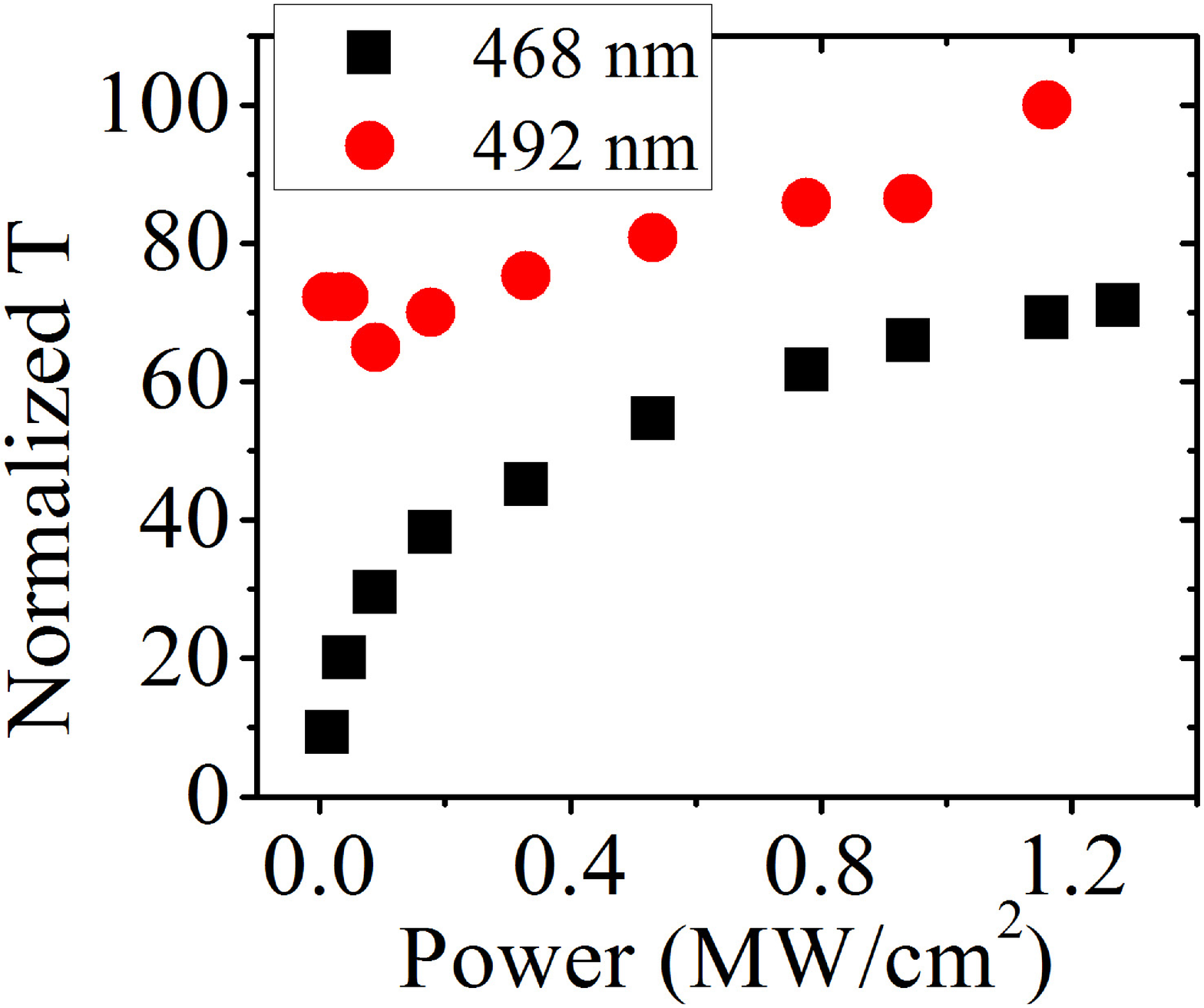}} \subfigure[]{\label{fig_4_b}%
\includegraphics[width=5.5cm, height=3.8cm]{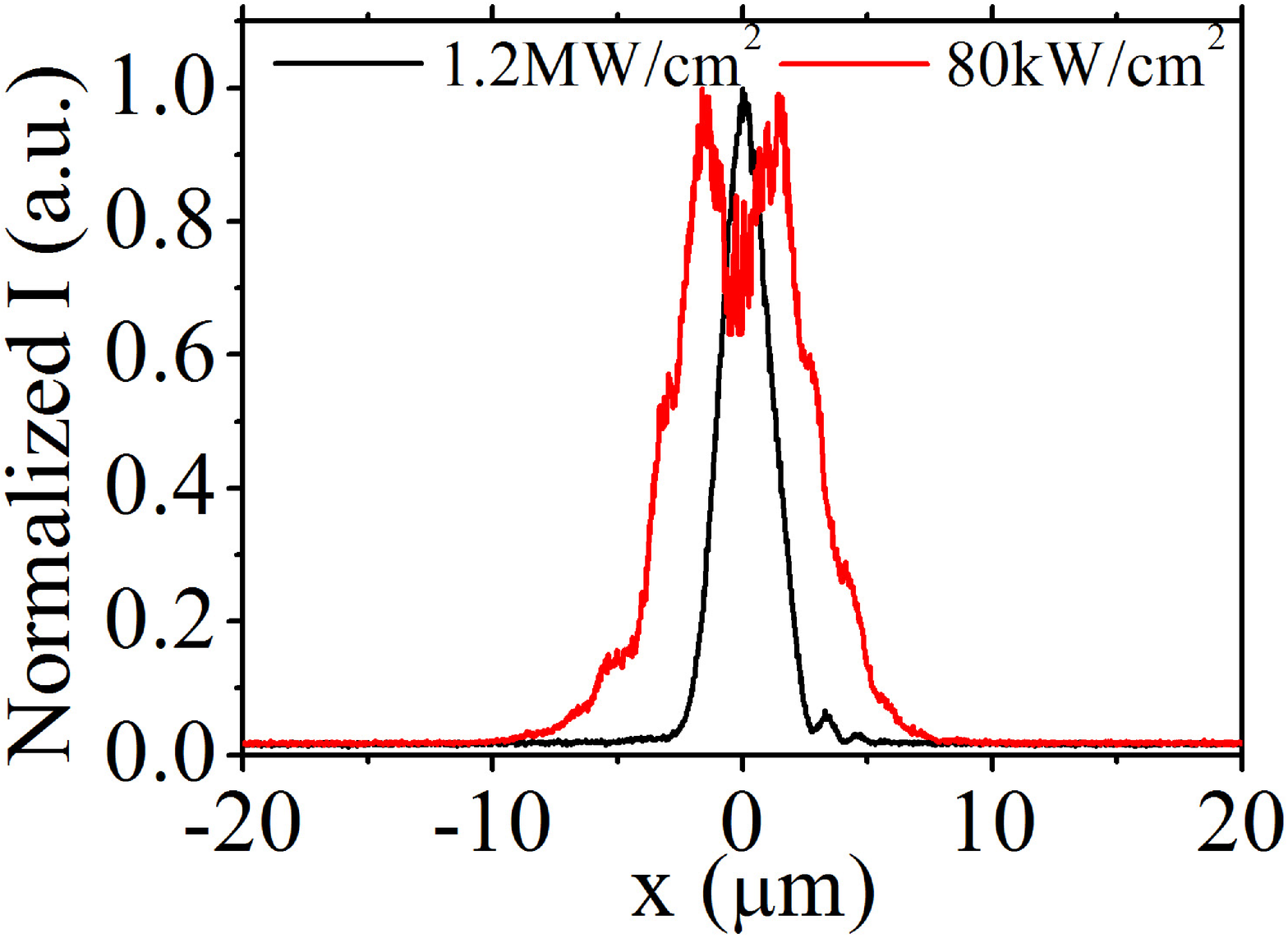}}
\caption{(Color online) (a) The relation between the transmission
coefficient (normalized to the maximum of the transmitted power in the
absence of the Bragg bandgap) at wavelengths $468$ nm and $492$ nm, and the
input intensity. (b) The saturation of the expansion of the output profile
at sufficiently high input intensities.}
\label{fig_4}
\end{figure}

\section{The theoretical model}
\indent The blue shift of the stopband indicates that the average refractive
index is reduced by the high-intensity pump, which may be considered as a
manifestation of the self-defocusing Kerr effect caused by the photo-induced
reorientation in the CLC. The couple-mode theory has been widely used for
the description of the nonlinear propagation in 1D photonic structures \cite%
{1,2,10}. As seen in Fig. 4(a), the bandgap shift increases with the pump
power, and then reaches saturation at the power exceeding $0.7$ MW/cm$^{2}$),
because of the absorption saturation by the dopant dye, as explained above.
Accordingly, the couple-mode equations can be introduced with the saturation
nonlinearity:
\begin{equation}
\pm i\frac{\partial E_{\pm }^{CP}}{\partial z}+F\frac{\partial ^{2}E_{\pm
}^{CP}}{\partial x^{2}}+\delta E_{\pm }^{CP}-\gamma \frac{|E_{\pm
}^{CP}|^{2}+|E_{\mp }^{CP}|^{2}}{1+c(|E_{\pm }^{CP}|^{2}+|E_{\mp }^{CP}|^{2})%
}E_{\pm }^{CP}+\kappa E_{\mp }^{CP}=0,  \label{CM}
\end{equation}%
where $|E_{\pm }^{CP}|^{2}$ denotes the power density of the forward and
backward circular-polarized (CP) waves, $z$ and $x$ are the propagation
distance and transverse coordinate, respectively, $\gamma =kn_{2}$ is the
nonlinearity strength,which can be changed by the external electric field,
with $k$ being the wavenumber and $n_{2}$ the material nonlinear
coefficient, $c$ is the saturation coefficient with the same dimension as $%
n_{2}$, $\kappa =\pi dn/\lambda _{B}$ is the effective reflectivity, where $%
dn$ is the amplitude of the refractive-index modulation, and $\delta =2\pi
(n_{\mathrm{eff}}\lambda ^{-1}-\lambda _{B}^{-1})$ measures the detuning of
the input from the Bragg wavelength, which is determined by the CLC period,
and can be altered by temperature. Factor $F=1/(2k)$ is the Fresnel
diffraction strength.

To analyze the spatial-propagation dynamics of light, we performed numerical
simulations of Eqs. (\ref{CM}) with the input wavelength located at the red
edge of the stopband with the following natural boundary conditions (b.c.):
\begin{equation}
E_{+}^{CP}(z=0,x)=A_{0}\mathrm{sech}(x/w),~E_{-}^{CP}(z=L,x)=0,  \label{bc}
\end{equation}%
where $A_{0}$ represents the amplitude of input field, $w$ is the beam's
width, and $L$ is the length of the sample. Equations (\ref{CM}) with b.c. (%
\ref{bc}) were solved by means of the finite-difference method. In the
simulations, the coordinates and field amplitude were scaled, so as to set
the span of the transverse coordinate to be $-1<x<+1,$ and the propagation
distance varying in the interval of $0<z<5$. The coupling coefficient is
fixed as $\kappa =0.7$, and detuning $\delta =-0.65$ (which is located at
the edge of the stopband). The other parameters were set as $\gamma =1.5$, $%
c=1$, $w=0.01$, and $F=4\times 10^{-5}$.

%
\begin{figure}[h]
\centering{\includegraphics[width=8 cm, height=6cm]{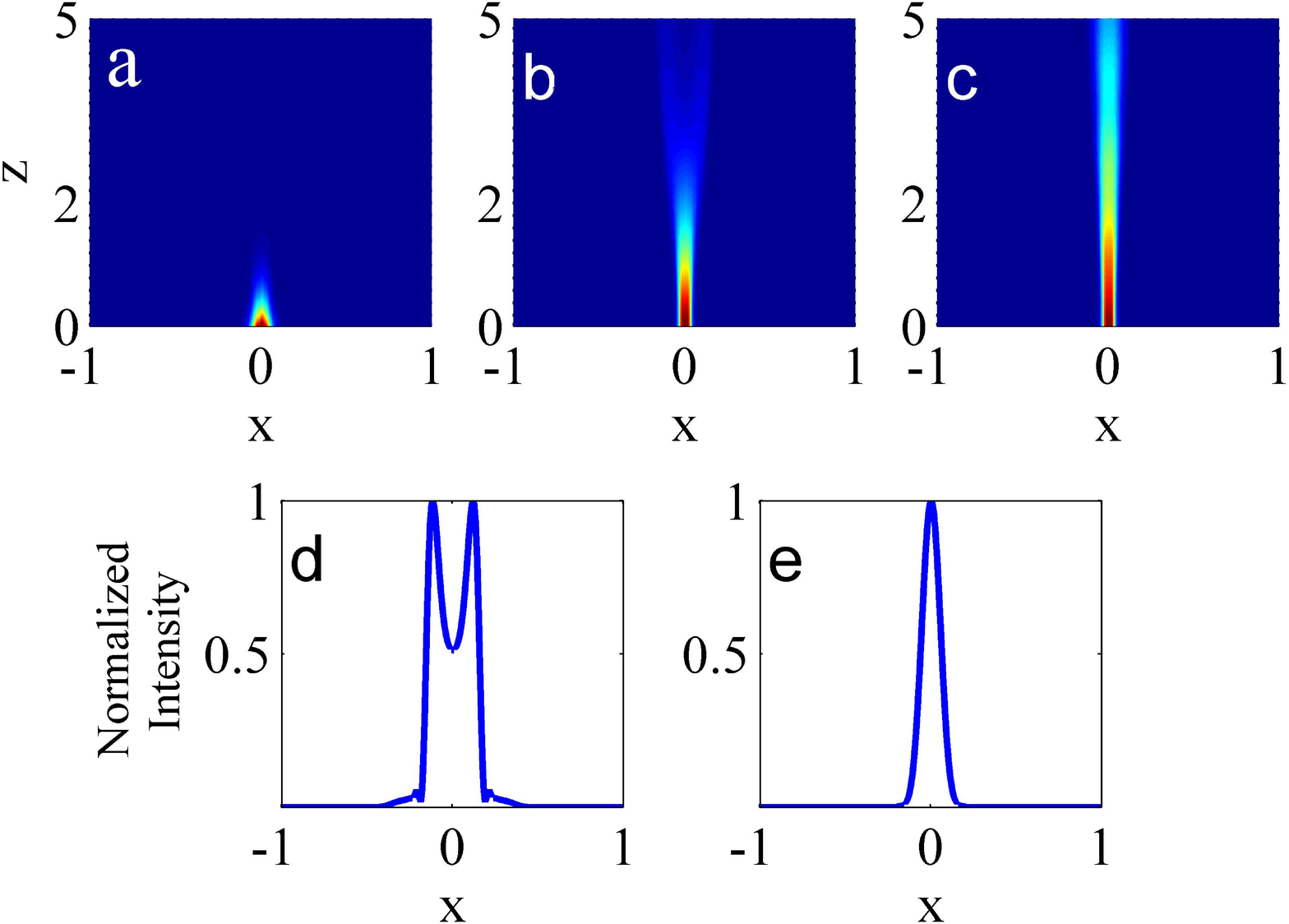}}
\caption{(Color online) Theoretical results for the propagation in the CLC,
showing that, in panel (a), the low- intensity beam cannot penetrate the
sample due to the reflection by the stopband. As the intensity increases,
the beam penetrates the sample, featuring diffraction due to the
self-defocusing nonlinearity. The respective propagation picture and the
beam cross section are shown in panels (b) and (d). As the intensity
increases further, the effective normal diffraction is established, see the
propagation picture and the corresponding beam cross section shown in panels
(c) and (e).}
\label{fig_5}
\end{figure}
\indent Typical examples of the simulated propagation are displayed in Fig.
5. At low input intensities, the input light is Bragg-reflected backward. As
the input intensity increases, the self-defocusing nonlinearity gives rise
to the blue shift of the stopband, which allows the input beam to propagate
through the sample with strong divergence. When the intensity increases
further, the divergence of the beam gets suppressed due to the saturation of
the nonlinearity. Finally, the self-defocusing is totally suppressed due to
the saturation, and the beam propagates effectively featuring the normal
linear diffraction. Detailed comparison (shown in Fig.5(b)) demonstrates
that the propagation regimes predicted by the simulations accurately match
the experimental observations.The normalized intensity in Fig. 5(e) is about 15 times larger than the normalized intensity in Fig. 5(d). The width at half maximum of Fig. 5(d) is 3 times wider than Fig. 5(e). This results are well in agreement with the experimental results.

\section{Conclusion}
\indent The subject of this work is the nonlinear light propagation in doped
samples of CLC. By using the dye-doped CLC, the nonlinear threshold and
nonlinearity length are substantially reduced, allowing one to achieve
strong modulation of the intensity and spatial distribution under low input
powers in a compact micro-device. The propagation of light with the
wavelength located near the edge of the stopband is studied experimentally
and simulated numerically. Due to the self-defocusing nonlinear effect in
the CLC, the intrinsic stopband can be shifted, leading to a substantial
enhancement of the transmission and strong divergence of the beam. As the
input power keeps increasing, the self-defocusing-induced divergence of the
beam is suppressed, due to the saturation of the nonlinearity. The system of
couple-mode equations was used to model the propagation of the light beams
in the CLC. Results of the simulations match the experimental observations
well. The setting introduced here may find applications to the design of
optical switching, power limiters, beam shapers, and optical buffering.

\ack{ The author thanks Professor I. Janossy for the useful
discussions.This project is supported by National Basic Research Program of
China (2012CB921904), and by the Chinese National Natural Science Foundation
(11074054). The work of Y. Li was supported by the German-Israel Foundation
through grant No. I-1024-2.7/2009, and by a postdoctoral fellowship from the
Tel Aviv University.}

\section*{References}
\begin{thebibliography}{99}
\bibitem{1} R. Slusher and B.
J. Eggleton,2003,\emph{Nonlinear Photonic Crystals}, (Berlin,Springer-Verlag).

\bibitem{2} Y. Kivshar and G. Agrawal, 2003,\emph{Optical Solitons: From Fibers
to Photonic Crystals} (London,Academic).

\bibitem{3} B. A. Malomed, D. Mihalache, F. Wise, and L. Torner,2005, J. Opt. B: Quant. Semicl. Opt. \textbf{7%
}, R53.

\bibitem{4} F. Lederer, G. I. Stegeman, D. N. Christodoulides, G. Assanto,
M. Segev, and Y. Silberberg,2008, Phys. Rep.
\textbf{463}, 1.

\bibitem{5} Y. V. Kartashov, V. A. Vysloukh, and L. Torner,2009, Progr. Opt. \textbf{52}, 63.

\bibitem{6} Y. V. Kartashov, B. A. Malomed, and L. Torner, 2011, Rev. Mod. Phys. \textbf{83}, 247.

\bibitem{15} M. Scalora, J. P. Dowling, C. M. Bowden, and M. J. Bloemer,1994
 Phys. Rev. Lett. \textbf{73}, 1368-1371.

\bibitem{16} S. Larochelle, Y. Hibino, V. Mizrahi, and G. I. Stegeman,1990,Electron. Lett. \textbf{26}, 1459-1460.

\bibitem{27} N. D. Sankey, D. F. Prelewitz, and T. G. Brown,1992, Appl. Phys. Lett.
\textbf{60}, 1427-1429.

\bibitem{28} P. Millar, R. M. De La Rue, T. F. Krauss, J. S. Aitchison, N.
G. R. Broderick, and D. J. Richardson, 1999, Opt. Lett. \textbf{24}, 685-687.

\bibitem{7} H. G. Winful, J. H. Marburger, and E. Garmire, 1979, Appl. Phys.
Lett. \textbf{35}, 379.

\bibitem{8} A. V. Andreev, A. V. Balakin, I. A. Ozheredov, A. P. Shkurinov,
P. Masselin, G. Mouret, D. Boucher, 2001, Phys. Rev. E. \textbf{63}%
, 016602.

\bibitem{9} F. Schreier and O. Bryngdahl, 2000, Opt. Commun. \textbf{185}, 227.

\bibitem{10} C. M. de Sterke, and J. E. Sipe,1994, Progr. Opt.
\textbf{33}, 205.

\bibitem{11} B. J. Eggleton, R. E. Slusher, C. M. de Sterke, P. A. Krug and
J. E. Sipe, 1996, Phys. Rev. Lett. \textbf{76}, 1627.

\bibitem{12} J. T. Mok, C. M. de Sterke, I. C. M. Littler, and B. J.
Eggleton, 2006, Nature Phys.
\textbf{2}, 775.

\bibitem{13} J. Y. Zhou, H. G. Shao, J. Zhao, and K. S. Wong, 2005, Opt.
Lett. \textbf{30}, 1560-1562.

\bibitem{13b} S. Fu, Y. Liu, Y. Li, L. Song, J. Li, B. A. Malomed, and J.
Zhou, 2013, Opt. Lett. \textbf{23}, 5047-5050.

\bibitem{14} J. P. Prineas, J. Y. Zhou, J. Kuhl, H. M. Gibbs, G. Khitrova,
S. W. Koch, and A. Knorr, 2002,
Appl. Phys. Lett. \textbf{81}, 4332-4334.

\bibitem{Silberberg} D. Mandelik, R. Morandotti, J. S. Aitchison, and Y.
Silberberg, 2004, Phys. Rev. Lett. \textbf{92}%
, 093904.

\bibitem{Feng} J. Feng, 1993, Opt. Lett. \textbf{18}, 1302-1304.

\bibitem{Nabiev} R. F. Nabiev, P. Yeh, and D. Botez, 1993, Opt. Lett. \textbf{18}, 1612-1614.

\bibitem{Mayer} J. Sch\"{o}llmann, R. Scheibenzuber, A. S. Kovalev, and A.
P. Mayer,1999, Phys. Rev. E. \textbf{59}, 4618-4629.

\bibitem{Atai} J. Atai and B. A. Malomed, 2001, Phys. Lett. A. \textbf{284}, 247-252.

\bibitem{Sukhorukov} A. A. Sukhorukov and Y. S. Kivshar, 2002, J. Opt. Soc. Am.
B. \textbf{19}, 772-781.

\bibitem{Ostrovskaya} B. A. Malomed, T. Mayteevarunyoo, E. A. Ostrovskaya,
and Y. S. Kivshar, 2005, Phys. Rev. E. \textbf{71}, 056616.

\bibitem{DeIre} A. Ciattoni, C. Rizza, E. DelRe, and E. Palange,
2007, Phys. Rev. Lett. \textbf{98},
043901.

\bibitem{17} N. Tamaoki, 2001, Advanced Materials. \textbf{13}, 1135.

\bibitem{J} I. Janossy, 1994, Phys. Rev. E. \textbf{49}%
, 2957.

\bibitem{YRShen} T. V. Truong, L. Xu, and Y. R. Shen, 2005, Phys. Rev. E. \textbf{72}, 051709.

\bibitem{M} L. Narducci, 2002, Liquid Crystals today,
\textbf{11}, 101002.

\bibitem{18} M. Li, P.-Q. Zhang, J. Guo, X.-S. Xie, Y.-K. Liu, B. Liang,
J.-Y. Zhou, and Y. Xiang, 2008, Chinese Phys. Lett.
\textbf{25}, 108.

\bibitem{19} X. Ying, M. Li, L. Tao, L. Jie, and J. Y. Zhou,
2007, Appl. Phys. A: Materials Science \&
Processing. \textbf{86}, 207 (2007).

\bibitem{20} Y. Xiang, T. Li, and J. Lin, 2006, Physics
Letters A. \textbf{357}, 159.

\bibitem{21} H.-C. Yeh, 2011, Opt.
Express. \textbf{19}, 5500.

\bibitem{22} J. Hwang, N. Y. Ha, H. J. Chang, B. Park, and J. W. Wu,
2004, Opt. Lett. \textbf{29}, 2644-2646.

\bibitem{apex} Tatsunosuke Matsui and Masahiro Kitaguchi, 2010, Appl. Phys. Express, \textbf{3}, 061701.

\bibitem{b1} V. A. Belyakov, V. E. Dmitrienko and V. P. Orlov,1979, Sov. Phys. Usp. \textbf{22} 64
doi:10.1070/PU1979v022n02ABEH005417

\bibitem{c1} G. S. Chilaya, 2006, Crystallogr. Rep.
\textbf{51}, S108.

\bibitem{23} F. A. Sala and M. A. Karpierz, 2012, Mol. Cryst. Liq. Cryst.
\textbf{558}, 176.

\bibitem{Fra}  A. Fratalocchi et al., Opt. Express \textbf{13}, 1808 (2005)

\bibitem{Dwei1}  D. Wei et al., ¡°Two-wave mixing in chiral dye-doped nematic liquid crystals¡± Optics Letters  \textbf{37},
734, (2012).

\bibitem{Dwei2}D. Wei, U. Bortolozzo, J. P. Huignard, and S. Residori, Slow and stored light by photo-isomerization induced transparency in dye doped chiral nematics,  Opt. Express, 21, 19544 (2013)

\bibitem{liyan} Liyan Song, Shenhe Fu, Yikun Liu, Jianying Zhou, Vladimir G.
Chigrinov, and Iam Choon Khoo,2013, Opt. Lett, \textbf{38},
5040.

\bibitem{KhooIC1} I. C. Khoo, Hong Li, Yu Liang, 1993, IEEE JOURNAL OF QUANTUM ELECTRONICS, \textbf{29}, 1444.

\bibitem{KhooIC2} I.C.Khoo, 1995,"Liquid Crystals: Physical Properties and
Nonlinear Optical Phenomena", (New York, Wiley).

\bibitem{lyK} P. Klysubun and G. Indebetouw, 2002,``Transient and steady state
photorefractive responses in dye-doped nematic liquid crystal cells," J.
Appl. Phys. \textbf{91}, 897.
\end{thebibliography}
\end{document}